%% file: main.tex
\documentclass[conference]{IEEEtran}
\usepackage{amsmath, mathrsfs,balance}
\usepackage{bm}

\usepackage{graphicx}
\usepackage{color}
\usepackage{tikz}
\usepackage{pgfplots}
\pgfplotsset{compat=newest}
\usetikzlibrary{patterns}
\usetikzlibrary{positioning}
\usetikzlibrary{datavisualization}
\usetikzlibrary{datavisualization.formats.functions}
\usetikzlibrary{backgrounds}
\usetikzlibrary{shapes,snakes}

\makeatletter
\def\maketag@@@#1{\hbox{\m@th\normalfont\normalsize#1}}
\makeatother

\usepackage[none]{hyphenat}

\newcommand{\subparagraph}{}
\usepackage[compact]{titlesec}
\titlespacing*{\section}{2pt}{1\baselineskip}{0.9\baselineskip}

\allowdisplaybreaks

\usepackage{amsfonts}
\usepackage{amssymb}
\usepackage{color}
\usepackage{multirow}
\usepackage{graphicx}

\usepackage{caption}
\usepackage{subcaption}
\usepackage[numbers,sort&compress]{natbib}
\usepackage{balance}

\input{symbols.tex}

\input{macros.tex}

\def\snrbef{{\mathsf{SNR}_\text{BBF}}}

\usepackage{tikz}
\usepackage{pgfplots}
\pgfplotsset{compat=newest}
\usetikzlibrary{patterns}
\usetikzlibrary{positioning}
\usetikzlibrary{datavisualization}
\usetikzlibrary{datavisualization.formats.functions}
\usetikzlibrary{backgrounds}
\usetikzlibrary{shapes,snakes}
\usepackage{textcomp}
\usetikzlibrary{automata}

\allowdisplaybreaks

\usepackage{algorithm}
\usepackage{algpseudocode}
\usepackage{pifont}
\usepackage{varwidth}

\def\one{{\bf 1}}

\usepackage{multicol,lipsum}
\usepackage{array}

\usepackage{lscape}

\def\herm{{\sfH}}

\def\ptot{{P_{\ttt \tto \ttt}}}

\def\cg{{\clC\clN}}

\usepackage{color}
\newcommand{\figref}[1]{Fig.~\ref{#1}}
\usetikzlibrary{decorations.markings}

\IEEEoverridecommandlockouts

\begin{document}

\sloppy

\title{Fully-Connected vs. Sub-Connected Hybrid Precoding Architectures for mmWave MU-MIMO
\thanks{X. Song is sponsored by the China Scholarship Council (201604910530). This work was funded by the European Union's Horizon 2020 research and innovation programme under grant agreement No. 779305 (SERENA).}}
\author{\IEEEauthorblockN{Xiaoshen Song, Thomas K\"uhne,   and Giuseppe Caire}
\IEEEauthorblockA{Communications and Information Theory Chair, Technische Universit\"at Berlin, Germany}
}

\maketitle

\begin{abstract}
Hybrid digital analog (HDA) beamforming has attracted considerable attention in practical implementation of millimeter wave (mmWave) multiuser multiple-input multiple-output (MU-MIMO) systems
due to its low power consumption with respect to its digital baseband counterpart. The implementation cost, performance, and power efficiency of HDA beamforming depends on the level of connectivity
and reconfigurability of the analog beamforming network. In this paper, we investigate the performance of two typical architectures for HDA MU-MIMO, i.e., the fully-connected (FC) architecture where each RF antenna port is connected to all antenna elements of the array, and the one-stream-per-subarray (OSPS) architecture where the RF antenna ports are connected to disjoint subarrays. 
We jointly consider the initial beam acquisition phase and data communication phase, such that the latter takes place by using the beam direction information obtained in the former phase. 
For each phase, we propose our own BA and precoding schemes that outperform the counterparts in the literature. 
We also evaluate the power efficiency of the two HDA architectures taking into account the practical hardware impairments, e.g., the power dissipation at different hardware components 
as well as the potential power backoff under typical power amplifier (PA) constraints. 
Numerical results show that the two architectures achieve similar sum spectral efficiency, but the OSPS architecture outperforms the FC 
case in terms of hardware complexity and power efficiency, only at the cost of a slightly longer time of initial beam acquisition. 
\end{abstract}	


\section{Introduction}\label{introduction}

Millimeter wave (mmWave)  multiuser multiple-input multiple-output (MU-MIMO) with large antenna arrays has been considered as a promising solution to meet the ever increasing data traffic for further $5$G wireless communications \cite{molisch2017hybrid}. Considering the unaffordable hardware cost of conventional full-digital baseband precoding, 
a combination of both digital and analog precoding, using a reduced number of RF chains, known as hybrid digital analog (HDA) structure, 
has been widely considered \cite{molisch2017hybrid}.

\begin{figure}[t]
	\centering
	\includegraphics[width=7.5cm]{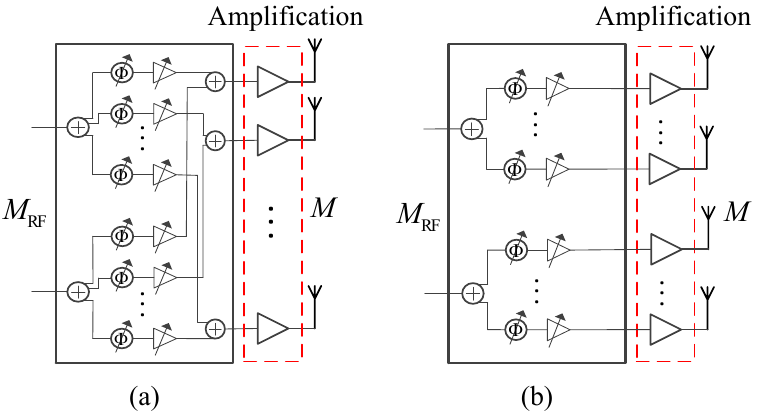}
	\caption{\small Hybrid transmitter architectures with (a) fully-connected (FC), and (b) sub-connected with one-stream-per-subarray (OSPS).}
	\label{TX}
\end{figure}

A large number of works is dedicated to the optimization and performance characterization of  HDA MU-MIMO architectures 
(e.g., see \cite{molisch2017hybrid,AngLi2017,JingboSubFull2018,CaoPerAntennaPower2018} and references therein).  However, we observe that the current literature has some significant shortcomings. 
In particular 1) Many works consider only phase-shift  control at the analog precoders \cite{AngLi2017,JingboSubFull2018,CaoPerAntennaPower2018}. This may somewhat reduce the hardware complexity, however, the signaling freedom is also drastically reduced. Also, it has been widely demonstrated in practical implementations that simultaneous amplitude and phase control as in \figref{TX} is feasible at mmWaves
with a low complexity and cost \cite{Castellanos2018}.  2) Many works ignore important  hardware impairments \cite{AngLi2017} such as 
the power dissipation and non-linear distortion of the  power amplifiers (PAs),  which have an important effect on the signal processing and should not be neglected. 
3) An overly large number of works investigate only the data communication phase and assume full  channel state information (CSI) \cite{AngLi2017,JingboSubFull2018,CaoPerAntennaPower2018}, i.e., 
they explicitly or implicitly assume that the precoder can be optimized using the channel coefficients seen at each antenna element, as if the signal from each antenna could be individually acquired.
In contrast, it is known that due to the severe isotropic pathloss, mmWave communication requires an initial acquisition phase (we refer here as  beam alignment (BA)) 
in order to find the strongest narrow beam pair connecting each  user equipment (UE) with the  base station (BS). 
Moreover, since the number of RF chains in a HDA architecture (see Fig.~\ref{TX}) is much smaller than the number of antenna elements, it is impossible 
to obtain at once all the channel coefficients from one round of training signals, as commonly done with digital baseband schemes. Hence, 
only the effective channel {\em from TX to RX antenna ports} along the beams acquired in the BA phase can be probed and measured.

In this paper, beyond the above fundamental limitations in the literature, we evaluate the performance of two typical transmitter architectures. 
On one hand, a  fully-connected (FC) architecture where each RF antenna port is connected to all antenna elements of the array (\figref{TX}$\,$(a)). 
At the other extreme, a  one-stream-per-subarray (OSPS) architecture where the RF antenna ports are connected to disjoint subarrays (\figref{TX}$\,$(b)). 
We jointly consider the initial BA phase, data communication, and practical hardware impairments. In particular, we propose our own BA/precoding scheme and evaluation model in order to 
provide a useful analysis and comparison framework for mmWave system design.

\section{Channel Model}\label{SystemModel}

We consider a system formed by a BS equipped with a  uniform linear array (ULA) 
with $M$ antennas and $M_{\text{RF}}$ RF antenna ports,\footnote{We use the term antenna port (3GPP terminology) to indicate 
the output (at the TX side) or the input (at the RX side) of individual RF modulation/demodulation chains, including A/D conversion.}
serving simultaneously  $K = M_{\text{RF}}$ UEs, each of which is also equipped with a ULA with $N$ antennas and $N_{\text{RF}}$ RF antenna ports. 
The propagation channel between the BS and the $k$-th UE, $k\in[K]$, consists of $L_k\ll \max\{M,N\}$ multi-path components, 
where the baseband equivalent impulse response of the channel at time slot $s$ reads
\begin{align}\label{ch_mod_disc_mp}
\sfH_{k,s}(t,\tau)&=\sum_{l=1}^{L_k} \rho_{k,s,l} e^{j2\pi \nu_{k,l}t}\bfa_{\text{R}}(\phi_{k,l}) \bfa_{\text{T}}(\theta_{k,l})^\herm \delta(\tau-\tau_{k,l})\nonumber\\
&=\sum_{l=1}^{L_k}\sfH_{k,s,l}(t)\delta(\tau-\tau_{k,l}),
\end{align}
where $\sfH_{k,s,l}(t) := \rho_{k,s,l} e^{j2\pi \nu_{k,l}t} \bfa_{\text{R}}(\phi_{k,l}) \bfa_{\text{T}}(\theta_{k,l})^\herm$, $(\phi_{k,l}, \theta_{k,l}, \tau_{k,l}, \nu_{k,l})$ denote the  angle of arrival (AoA), angle of departure (AoD), delay, and Doppler shift 
of the $l$-th component, and $\delta(\cdot)$ denotes the Dirac delta function.
The vectors $ \bfa_{\text{T}}(\theta_{k,l})\in \bC^{D}$  and $ \bfa_{\text{R}}(\phi_{k,l})\in \bC^{N}$ are the array response vectors of the BS and UE at AoD $\theta_{k,l}$ and AoA $\phi_{k,l}$ respectively, with elements given by
\begin{subequations}  \label{array-resp}
	\begin{align}
	[\bfa_{\text{T}}(\theta)]_d&=e^{j (d-1)\pi \sin(\theta)}, \ d \in[D], \label{a_resp_BS}\\
	[\bfa_{\text{R}}(\phi)]_n&=e^{j (n-1)\pi \sin(\phi)}, \ n\in[N],\label{a_resp_UE}
	\end{align}
\end{subequations}
where $D=M$ for \figref{TX}(a) and $D=\frac{M}{M_{\text{RF}}}$ for \figref{TX}(b). Here we assume that the spacing of the ULA antennas equals to the half of the wavelength. We adopt a block fading model, i.e., the channel gains $\rho_{k,s,l}$ remain invariant over the
channel \textit{coherence time} $\Delta t_c$ but change i.i.d. randomly across different $\Delta t_c$.
Since each scatterer in practice is a superposition of many smaller components that have (roughly) the same AoA-AoD and delay, we assume a general 
Rice fading model given by
\begin{align}\label{rice_fading}
\rho_{k,s,l}\sim \sqrt{\gamma_{k,l}} \left(\sqrt{\frac{\eta_{k,l}}{1+\eta_{k,l}}}+\frac{1}{\sqrt{1+\eta_{k,l}}}\check{\rho}_{k,s,l}\right),
\end{align}
where $\gamma_{k,l}$ denotes the overall multi-path component strength, 
$\eta_{k,l}\in[0,\infty)$ indicates the strength ratio between the line-of-sight (LOS) and the  non-LOS (NLOS) components, 
and $\check{\rho}_{k,s,l} \sim \cg(0, 1)$ is a zero-mean unit-variance complex Gaussian random variable. 
In particular, $\eta_{k,l}\to \infty$ indicates a pure LOS path while  $\eta_{k,l}=0$ indicates a pure NLOS path, affected by standard Rayleigh fading.

Following the {\em beamspace representation} as in \cite{sxs2018TimeJour}, we obtain an approximate finite-dimensional representation of the channel response \eqref{ch_mod_disc_mp} 
with respect to the discrete dictionary in the AoA-AoD (beam) domain defined by the quantized angles
\begin{subequations} \label{theta-phi}
	\begin{align}\label{gridtheta}
	\Phi&:=\{\check{\phi}: (1+\sin(\check{\phi}))/2=\frac{n-1}{N}, \, n \in [N]\},\\
	\Theta&:=\{\check{\theta}: (1+\sin(\check{\theta}))/2=\frac{d-1}{M}, d\in [D]\},
	\end{align}
\end{subequations}
with corresponding array response vectors $\clA_{\text{R}}:=\{\bfa_{\text{R}}(\check{\phi}): \check{\phi} \in \Phi\}$ and $\clA_{\text{T}}:=\{\bfa_{\text{T}}(\check{\theta}): \check{\theta} \in \Theta\}$. 
For ULAs as considered in this paper, the dictionaries $\clA_{\text{R}}$ and $\clA_{\text{T}}$, after suitable normalization, yield to the  discrete Fourier transform (DFT) matrices 
$\bfF_{N}\in \bC^{N\times N}$ and $\bfF_{D}\in \bC^{D\times M}$ with elements
\begin{subequations}
	\begin{align}
	[\bfF_{N}]_{n,n'}&=\frac{1}{\sqrt{N}}e^{j2\pi (n-1)(\frac{n'-1}{N}-\frac{1}{2})}, n,n'\in[N],\\
	[\bfF_{D}]_{d,d'}&=\frac{1}{\sqrt{M}}e^{j2\pi (d-1)(\frac{d'-1}{M}-\frac{1}{2})}, d\in[D],d'\in[M].
	\end{align}
\end{subequations}
Consequently, the beam-domain channel representation reads 
\begin{align}\label{beamspacechannel}
\check{\sfH}_{k,s}(t,\tau)\! =\! \bfF_{N}^\herm\sfH_{k,s}(t,\tau)\bfF_{D}\! =\! \sum_{l=1}^{L_k}\check{\sfH}_{k,s,l}(t)\delta(\tau-\tau_{k,l}),
\end{align}
where $\check{\sfH}_{k,s,l}(t) := \bfF_{N}^\herm\sfH_{k,s,l}(t)\bfF_{D}$. It is well-known (e.g., see \cite{sxsBA2017} and references therein) that, 
as $M$ and $N$ increase, the DFT basis provides a very sparse channel representation.

\section{Beam Acquisition and Data Transmission} 

\begin{figure}[t]
	\centering
	\includegraphics[width=7.5cm]{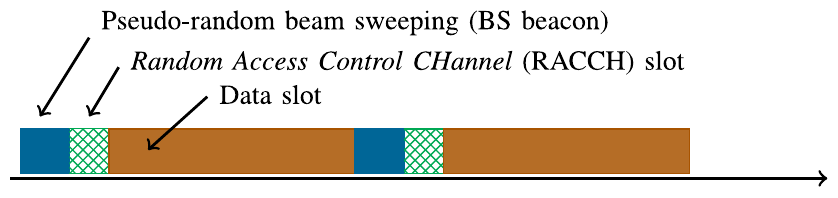}
	\caption{{\small Illustration of the frame structure in the underlying system.}}
	\label{2_frame_timeline}
\end{figure}
\figref{2_frame_timeline} illustrates the considered frame structure which consists of three parts \cite{sxs2018TimeJour}: the beacon slot, the  random access control channel (RACCH) slot, and the data slot. As illustrated in our previous work \cite{sxs2018TimeJour}, in the initial acquisition phase the measurements are collected at the UEs from downlink beacon slots broadcasted by the BS.	
Each UE selects its strongest AoA as the beamforming direction for possible data transmission. During the RACCH slot, the BS stays in listening mode such that each UE sends a beamformed packet to the BS. This packet contains basic information such as the UE ID and the beam indices of the selected AoDs. The BS responds with an acknowledgment data packet in the data subslot of a next frame. 
From this moment on the BS and the UE are connected in the sense that, if the procedure is successful, they have achieved BA. In other words, 
they can communicate by aligning their beams along a  mutlipath component with AoA-AoD $(\phi_{k,l}, \theta_{k,l})$ and strong coefficient $\rho_{k,l}$. 

The details of the BA algorithm and its performance are given in \cite{sxsBA2017} for a frequency-domain based variant of the problem, 
and in \cite{sxs2018TimeJour} for a time-domain single-carrier variant of the problem. In both cases, we have shown that the proposed scheme aligns the beams along the strongest multipath
component with probability that tends to 1 as the number of beacon slots (measurements) increases, and that the probability of collision or errors on the RACCH protocol information exchange is
negligible when the alignment directions have been correctly found. 

\section{System Analysis}

\subsection{Hardware Impairments}\label{hardwareImperfection}

We assume that each analog path has simultaneous amplitude and phase control as shown in \figref{TX}. Let $\tilde{\bfx}=[\tilde{x}_1,\cdots,\tilde{x}_M]\in\bC^M$ denote the beamformed signal\footnote{Here for notation simplicity, we neglected the signal time-domin index $t$.} given by
\begin{align}\label{beamformedX}
	\tilde{\bfx} = \sqrt{\alpha_{\text{com}}} \tilde{\bfU}\cdot\sqrt{\alpha_{\text{div}}} \bfx,
\end{align}
where $\bfx=[x_1,\cdots, x_{M_{\text{RF}}}]\in\bC^{M_{\text{RF}}}$ is the transmit complex symbol vector, with $\bE[|x_k|^2]=\epsilon$, $k\in[M_{\text{RF}}]$. $\alpha_{\text{div}}$ results from to the signal splitting with $\alpha_{\text{div}}=\frac{1}{M}$ for \figref{TX} (a) and $\alpha_{\text{div}}=\frac{M_\text{RF}}{M}$ for \figref{TX} (b). $\alpha_{\text{com}}$ models the power dissipation factor of the combiners corresponding to their S-parameters as in \cite{JingboSubFull2018} with $\alpha_{\text{com}} = \frac{1}{M_\text{RF}}$ for \figref{TX} (a) and $\alpha_{\text{com}} = 1$ for \figref{TX} (b). $\tilde{\bfU}\in\bC^{M\times M_{\text{RF}}}$ denotes the overall beamforming coefficients, given by
\begin{align}
	[\tilde{\bfu}_1, \tilde{\bfu}_2,\cdots, \tilde{\bfu}_{M_{\text{RF}}}] \quad\text{and} \quad \begin{bmatrix}
	\tilde{\bfu}_1 & 0 & ...& 0 \\
	0 & \tilde{\bfu}_2 &...& 0 \\
	\vdots & \vdots &\ddots & \vdots \\
	0 & 0 &...& \tilde{\bfu}_{M_{\text{RF}}} 
	\end{bmatrix}
\end{align}
for \figref{TX} (a) and \figref{TX} (b) respectively, where to meet the total input constraint, in \figref{TX} (a) we have $\tilde{\bfu}_k\in\bC^{M}$ with $\|\tilde{\bfu}_k\|^2=M$, whereas in \figref{TX} (b) we assume $\tilde{\bfu}_k\in\bC^{\frac{M}{M_{\text{RF}}}}$ with $\|\tilde{\bfu}_k\|^2=\frac{M}{M_{\text{RF}}}$, $k\in[M_{\text{RF}}]$. Based on \eqref{beamformedX}, the sum-power of the beamformed signal $\tilde{\bfx}$ can be written as
\begin{align}
\tilde{P} &= \bE[\tilde{\bfx}^\herm\tilde{\bfx}]=\alpha_{\text{com}}\alpha_{\text{div}}\cdot\bfx^\herm\tilde{\bfU}^\herm\tilde{\bfU}\bfx\nonumber\\
&=\alpha_{\text{com}}\alpha_{\text{div}}\cdot\trace\left(\bfx\bfx^\herm\tilde{\bfU}^\herm\tilde{\bfU}\right).
\end{align}
Consequently, the sum-power for the FC architecture of \figref{TX} (a) and for the OSPS architecture of \figref{TX} (b) reads 
$\tilde{P}_{\text{FC}} = \epsilon M_{\text{RF}} \frac{1}{M_{\text{RF}}}$ and $\tilde{P}_{\text{OSPS}}=\epsilon M_{\text{RF}}$, respectively. 
In order to compensate the additional combiner power dissipation in \figref{TX} (a), the transmitter should either boost the input signal as $M_{\text{RF}} \bfx$ (with low cost) or choose PAs with 
larger gain for the amplification stage (with higher cost). We consider the former approach because it is more cost effective. Further, we include the parameters $(\alpha_{\text{com}},\alpha_{\text{div}})$ into the column-wise normalized beamforming matrix $\widetilde{\bfU}$, such that we can 
write \eqref{beamformedX} as
\begin{align}
\tilde{\bfx} = \widetilde{\bfU} \bfx.
\end{align}

The beamformed signal then goes through the amplification stage, where at each antenna branch a PA amplifies the signal before transmission. We assume that the PAs in different antenna branches have
the same input-output relation. For any given antenna in the transmitter array, let $P_{\text{rad}}$ denote the radiated power of the antenna, and $P_{\text{cons}}$ denote the consumed power by the corresponding PA including both the radiated power and the dissipated power. Following the approach in \cite{Moghadam2018}, the  power consumed by the PA reads
\begin{align}
P_{\text{cons}} = \frac{\sqrt{P_{\text{max}}}}{\eta_{\text{max}}}\sqrt{P_{\text{rad}}},
\end{align}
where $P_{\text{max}}$ is the maximum output power of the PA with $P_{\text{rad}}\leq P_{\text{max}}$, and $\eta_{\text{max}}$ is the maximum efficiency of the PA. Considering that the PAs are often the predominant power consumption part, we define $\eta_{\text{eff}}$ given by
\begin{align}
\eta_{\text{eff}} = \frac{P_{\text{rad}}}{P_{\text{cons}}}
\end{align}
as the metric to effectively compare the power efficiency of the two transmitter architectures in \figref{TX}.

Due to the superposition of multiple beamforming vectors (particularly in the FC case) and the potentially high  peak to average power ratio (PAPR) of the time-domain transmit waveform $x_k$ (particularly with orthogonal frequency division multiplexing), the input power for some individual PA may exceed its saturation limit. This would result in non-linear distortion and even the collapse of the whole transmission. To compare the two transmitter architectures and ensure that all the underlying $M$ PAs simultaneously work in their linear range, we generally have two options:

{\em\textbf{ Option I:}} All transmitters utilize the same PA but apply a different input back-off $\alpha_{\text{off}}\in(0,1]$, such that the peak power of the radiated signal is smaller than $ P_{\text{max}}$. As a reference, we denote by $(P_{\text{rad},0},\eta_{\text{max},0})$ as the parameters of a reference PA under the reference precoding/beamforming strategy with a power backoff factor $\alpha_{\text{off},0}$ (as illustrated later in Section \ref{Numerical}). For different scenarios (with certain $\alpha_{\text{off}}$) the effective radiated power and the consumed power read $P_{\text{rad}} = \frac{\alpha_{\text{off}}}{\alpha_{\text{off},0}}P_{\text{rad},0}$, $P_{\text{cons}} = \frac{\sqrt{P_{\text{max},0}}}{\eta_{\text{max},0}}\sqrt{P_{\text{rad}}}$. The transmitter efficiency is given by
\begin{align}\label{eff1}
\eta_{\text{eff}} = \frac{P_{\text{rad}}}{P_{\text{cons}}} = \frac{\sqrt{P_{\text{rad}}}\cdot\eta_{\text{max},0}}{\sqrt{P_{\text{max},0}}}.
\end{align}

{\em\textbf{ Option II:}} We choose to deploy different PAs for different transmitter architectures, with a maximum output power given by $P_{\max}\! =\! \frac{\alpha_{\text{off},0}}{\alpha_{\text{off}}}P_{\text{max},0}$, where $\alpha_{\text{off}}$ has the same value as in {\em Option I}. Consequently, the effective radiated power and the consumed power of the underlying PA read $P_{\text{rad}} = P_{\text{rad},0}$, $P_{\text{cons}}\! = \!\frac{\sqrt{P_{\text{max},0}\cdot\alpha_{\text{off},0}/\alpha_{\text{off}}}}{\eta_{\text{max}}}\sqrt{P_{\text{rad}}}$. The transmitter efficiency is given by
\begin{align}
\eta_{\text{eff}} = \frac{P_{\text{rad}}}{P_{\text{cons}}} = \frac{\sqrt{P_{\text{rad}}}\cdot\eta_{\text{max}}}{\sqrt{P_{\text{max},0}\cdot \alpha_{\text{off},0}}}\cdot \sqrt{\alpha_{\text{off}}}.
\end{align}

Note that the characteristics $(P_{\text{max}}$ and $\eta_{\text{max}})$ of different PAs highly depend on the operation frequency, implementation, and technology. Aiming at illustrating how to apply the proposed analysis framework in practical system design, we will exemplify a set of PA parameters in Section \ref{Numerical} to evaluate the efficiency $\eta_{\text{eff}}$ of the two architectures in \figref{TX}. However, in the following derivations for the BA and data communication, otherwise stated, we will assume a  single-carrier (SC) modulation and a fixed total radiated power constraint denoted by $\ptot$, where all the underlying PAs work in their linear range with an identical scalar gain.

\subsection{Beam Alignment (BA) Phase}\label{beamAlignment}
As discussed in Section \ref{introduction}, communication at mmWaves requires narrow beams via large antenna array beamforming to overcome the severe signal attenuation. In this section, we provide a brief description of our recently proposed time-domain BA scheme and refer to \cite{sxs2018TimeJour} for more details.

In short, we assume that the BS broadcasts its pilot signals periodically over the beacon slots according to a pseudo-random beamforming codebook, which is known to all the UEs in the system. We assign a unique  Pseudo Noise (PN) sequence as the pilot signal to each RF chain at the BS such that different pilot streams are separable at the UE. Meanwhile, each UE independently collects its measurements to estimate its strong AoA-AoD combinations. Taking the $k$-th UE for example, let $\Gammam_k$ denote an all-zero $N\times M$ matrix with positive elements corresponding to the beam-domain second-order statistics of the channel coefficients, given by
\begin{align}\label{averageGamma}
[\Gammam_k]_{n,m} \propto \sum_{l=1}^{L_{k}}\bE\left[\left|[\check{\sfH}_{k,s,l}(t,\tau_{k,l})]_{n,m}\right|^2\right].
\end{align} 
Over $T$ beacon slots the UE obtains a total number of $M_{\text{RF}}N_{\text{RF}}T$ equations, which can be written in the form
\begin{align}\label{UE_equations}
\bfq_k=\bfB_k \cdot\vec{(\Gammam_k)} + \zeta(\ptot)\cdot \one + \bfw_k,
\end{align}
where $\bfq_k\in \bR^{M_{\text{RF}}N_{\text{RF}}T}$ consists of all the $M_{\text{RF}}N_{\text{RF}}T$ statistical power measurements, $\bfB_k\in \bR^{M_{\text{RF}}N_{\text{RF}}T\times MN}$ is uniquely defined by the pseudo-random beamforming codebook of the BS and the local beamforming codebook of the $k$-th UE, $\zeta(\ptot)$ denotes a constant whose value is a function of the total radiated power, and  $\bfw_k \in \bR^{M_{\text{RF}}N_{\text{RF}}T}$ denotes the residual measurement fluctuations. As discussed in \cite{sxs2018TimeJour}, with the non-negative constraint of $\Gammam_k$, a simple  Least Squares (LS)
\begin{align}\label{eq:NNLS}
\Gammam_k^\star=\argmin_{\Gammam_k \in \bR_+^{N\times M}} \|\bfB \cdot\vec{(\Gammam_k)} + \zeta(\ptot)\cdot \one - \bfq_k\|^2
\end{align}
is sufficient to impose the sparsity of the solution $\Gammam_k^\star$. We assume a success of the BA if the largest component in $\Gammam_k^\star$ coincides with the actual strongest path of the $k$-th UE. More details can be found in \cite{sxs2018TimeJour}.

\subsection{Data Communication Phase}
We assume that the BS simultaneously schedules $K= M_{\text{RF}}$ UEs which are selected by a simple directional scheduler \cite{yunyi2017scheduler}. Namely, the selected $K$ UEs have similar power profiles and their strongest AoDs in the downlink are at least $\Delta\theta_{\text{min}}$ away from each other. Consequently, the multi-user beamforming scheme at the BS allocates equal power across these UEs with potential multi-user interference cancellation \cite{yunyi2017scheduler}. Denoted by $\bfu_k$ as the normalized transmit beamforming vector for the $k$-th UE at the BS and $\bfv_k$ as the normalized receive beamforming vector at the $k$-th UE, with an effective radiated power $P_k=\frac{\ptot}{M_{\text{RF}}}$ to maintain the total radiated power constraint, the received signal at the $k$-th UE can be written as
\begin{align}\label{data_out1}
y_k(t)=&\bfv_{k}^\herm\sum_{k'=1}^{K}\sqrt{P_{k'}}\sfH_{k,s}(t,\tau)\circledast \left(\bfu_{k'} x_{k'}(t)\right)+z_k(t)\nonumber\\
=&\sqrt{P_{k}}\left(\bfv_{k}^\herm\sfH_{k,s}(t,\tau)\bfu_{k}\right)\circledast x_{k}(t)+z_k(t) \nonumber\\[4pt]
+& \sum_{k'\neq k}\sqrt{P_{k'}}\left(\bfv_{k}^\herm\sfH_{k,s}(t,\tau)\bfu_{k'}\right)\circledast x_{k'}(t)
\end{align}
where $f(t)\circledast g(t) = \int f(\tau)g(t-\tau)d\tau$ denotes the convolution operation. As we can see, the first term in \eqref{data_out1} corresponds to the desired signal at the $k$-th UE, whereas the last two terms correspond to the noise and interference, respectively. By substituting \eqref{ch_mod_disc_mp} into \eqref{data_out1}, the received signal reads
\begin{align}\label{data_out}
y_k(t)
&= \sum_{l=1}^{L_k} \sqrt{P_{k}} \bfv_{k}^\herm\sfH_{k,s,l}(t)\bfu_{k}x_{k}(t-\tau_{k,l})+z_k(t)\nonumber\\
&+\sum_{k'\neq k}\sum_{l=1}^{L_{k'}}  \sqrt{P_{k'}}\bfv_{k}^\herm\sfH_{k',s,l}(t)\bfu_{k'}x_{k'}(t-\tau_{k',l}) ,
\end{align}
where $x_k(t)$ denotes the unit-power transmit signal, $z_k(t)\sim \cg(0, N_0B)$ denotes the continuous-time complex  additive white Gaussian noise (AWGN) with a  power spectral density (PSD) of $N_0$ Watt/Hz, and $B$ denotes the effective bandwidth. By treating the multi-user interference as noise at each UE the asymptotic spectral efficiency of the $k$-th UE is given by  
\begin{align}\label{SE}
R_k \!\!=\!\bE\!\!\left[\!\log_2\!\left(\!\!1\!+\!\frac{{P_{k}}|\sum_{l=1}^{L_k}\bfv_{k}^\herm \sfH_{k,s,l}(t)\bfu_{k}|^2}{|\!\!\sum\limits_{k'\neq k}\!\!\!\sum_{l=1}^{L_{k'}} \!\sqrt{P_{k'}}\bfv_{k}^\herm\sfH_{k'\!,s,l}(t)\bfu_{k'}|^2\!\!+\!|z_k(t)|^2}\!\!\right)\!\!\right],
\end{align}
and the sum spectral efficiency reads $R_{\text{sum}} =\sum_{k=1}^{K}R_k$.

We claim that the beamforming vectors corresponding to each UE in the data communication phase are based on the outcome of the BA in Section \ref{beamAlignment}. More precisely, assume that after a BA procedure the strongest component in $\Gammam_k^\star$ corresponds to the $l_k$-th multi-path component in $\check{\sfH}_{k,s}(t,\tau)$ between the BS and the $k$-th UE. To simplify the practical implementation, we assume that the $k$-th UE decodes its data along the estimated strongest direction, given by 
\begin{align}\label{receive_vector}
\bfv_{k} = \Fm_N \check{\bfv}_{k},
\end{align}
where $ \check{\bfv}_{k}\in\bC^N$ is an all-zero vector with a $1$ at the component corresponding to the AoA of the $l_k$-th scatterer. At the BS we assume that the BS communicates with the $k$-th UE along its top-$p$ beams with respect to the AoA given by $\bfv_{k}$, where $p\geq1$ in order to handle the potential mobility or blockage\footnote{Not to be confused with the case that both UE and BS point to multiple directions. In our scheme, the UE always points to a single strongest direction, whereas the BS points to $p$ directions with respect to each UE.}. Define $\bfU_k\in\bC^{D\times p}$, each column of which corresponds to one of the $p$ AoD beamforming directions, given by 
\begin{align}\label{RF_precoder}
\bfU_{k} &= [\bfu_{k,1}, \bfu_{k,2},\, ..., \bfu_{k,p}]\nonumber\\
&=\bfF_{D}\cdot[\check{\bfu}_{k,1}, \check{\bfu}_{k,2},\, ..., \check{\bfu}_{k,p}],
\end{align}
where $\check{\bfu}_{k,i}\in\bC^{M}$, $i\in[p]$, is an all-zero vector with a $1$ at the component corresponding to the $i$-th strongest AoD direction of the $k$-th UE, and where $D=M$ for the FC 
architecture, otherwise $D=\frac{M}{M_{\text{RF}}}$ for the OSPS case.

To formulate the hybrid precoding problem, we re-write everything in a matrix-multiplication format. Let $\bfx(t)=\diag(\sqrt{P_1}, \sqrt{P_2}, ..., \sqrt{P_K})\cdot [x_1(t), x_2(t), ..., x_K(t)]^{\transp}\in\bC^{K}$ denote the transmit signal vector and $\overline{\sfH}_s(t,\tau)$ denote the aggregated channel for all the $K$ UEs given by
\begin{align}
\overline{\sfH}_s(t,\tau) \!=\! \left[\sfH_{1,s}(t,\tau)^\transp, \sfH_{2,s}(t,\tau)^\transp, \cdots, \sfH_{K,s}(t,\tau)^\transp\right]^\transp\!,
\end{align}
where $\sfH_{k,s}(t,\tau)$, $k\in[K]$, is given in \eqref{ch_mod_disc_mp}. We define  $\bfV\in\bC^{NK\times K}$ as the receive beamforming matrix given by
\begin{align}
\bfV&= \diag(\bfv_{1}, \bfv_{2},\, ..., \bfv_{K})\nonumber\\
&=(\bfI_K \otimes \bfF_N)\cdot \diag(\check{\bfv}_{1}, \check{\bfv}_{2}, ..., \check{\bfv}_{K}),
\end{align}
where $\bfI_K$ denotes the $K\times K$ identity matrix, and $\otimes$ represents the Kronecker product. Let $\overline{\bfU}\in\bC^{D\times pK}$ denote the analog precoding vector support given by
\begin{align}\label{RF_support}
	\overline{\bfU} &= [\bfU_{1}, \bfU_{2}, ..., \bfU_{K}]\nonumber\\
	&=\bfF_D\cdot [\check{\bfu}_{1,1}, ..., \check{\bfu}_{1,p}, ..., \check{\bfu}_{K,1}..., \check{\bfu}_{K,p}],
\end{align}
and $\bfA_B = [\bfa_1,\bfa_2, ..., \bfa_K]\in\bC^{pK\times K}$ denote the baseband precoding matrix. The precoding matrix $\bfU\in\bC^{D\times K}$ at the BS can be written as
\begin{align}\label{precoder}
\bfU = [\bfu_1, \bfu_2, ..., \bfu_K] = \overline{\bfU}\cdot \bfA_B.
\end{align} 
To meet the total radiated power constraint the coefficients in \eqref{precoder} are normalized as $\|\bfu_k\|=\|\overline{\bfU}\cdot \bfa_k\|=1$. As a result, the receive signal $\bfy(t)=[y_1(t), y_2(t), ..., y_K(t)]^{\transp}\in\bC^{K}$ reads
\begin{align}\label{receive}
\bfy(t) &= \bfV^\herm \cdot \overline{\sfH}_s(t,\tau) \circledast \left(\bfU \cdot\bfx(t)\right)  +\bfz(t)\nonumber\\
& = \left(\bfV^\herm \cdot \overline{\sfH}_s(t,\tau) \cdot \overline{\bfU}\cdot \bfA_B\right)\circledast\bfx(t) +\bfz(t)\nonumber\\
& := \left(\widetilde{\sfH}_{s}(t,\tau)\cdot \bfA_B\right)\circledast\bfx(t)+\bfz(t),
\end{align} 
where $\bfz(t)\in\bC^{K}$ denotes the noise, and 
\begin{align}\label{EF_H}
\widetilde{\sfH}_{s}(t,\tau)=\bfV^\herm \cdot \overline{\sfH}_s(t,\tau) \cdot \overline{\bfU}
\end{align} 
represents the $K\times (p\,K)$-lower-dimensional effective channel. The effective channel can be easily estimated over $p\cdot K$ additional sub-slots in the uplink, on the condition that a successful BA procedure in Section \ref{beamAlignment} has been achieved\footnote{We use channel reciprocity and standard uplink orthogonal pilot transmission for the lower-dimensional effective channel estimation of $\widetilde{\sfH}_{s}(t,\tau)$.}.

%

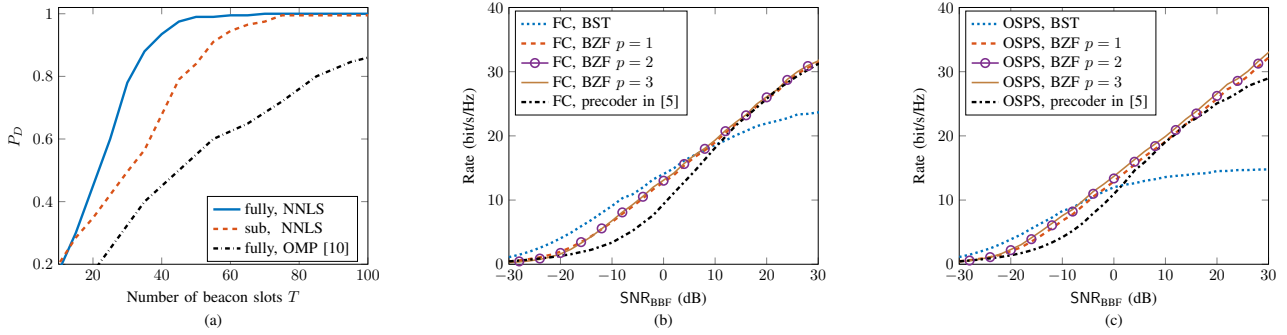
\begin{figure*}[t]
	\centering	
	\hspace{-0.8cm}
	\begin{subfigure}[b]{0.3\textwidth}
		\centering
		\scalebox{.6}{\input{1-BA-compare}}%
	\end{subfigure}
	\hspace{0.3cm}
	\begin{subfigure}[b]{0.3\textwidth}
		\centering
		\scalebox{.6}{\input{1-fully-rate-snr}}%
	\end{subfigure}
	\hspace{0.3cm}
	\begin{subfigure}[b]{0.3\textwidth}
		\centering
		\scalebox{.6}{\input{2-sub-rate-snr}}%
	\end{subfigure}
	\caption{{\small (a) Detection probability $P_D$ of different transmitter architectures vs. the training overhead, for the initial beam alignment phase with $\snrbef=-20$ dB. (b) The sum spectral efficiency of 
	the FC architecture  vs. increasing $\snrbef$, for the data communication phase with different precoding schemes. (c) The sum spectral efficiency of the OSPS architecture  vs. increasing $\snrbef$, for the data communication phase with different precoding schemes.}}
	\label{1-BA-SE}
\end{figure*}

\subsubsection{Beam Steering (BST) Scheme}
In the beam steering (BST) scheme, we assume that the BS simply steers $K$ data streams towards the $K$ strongest AoDs, i.e., we have $p=1$ in \eqref{RF_precoder}. As illustrated in Section \ref{SystemModel}, the underlying beam indices are estimated and fed back from the corresponding $K$ UEs. More preciously, assume that after a BA procedure as in Section \ref{beamAlignment}, the strongest component (top first) in $\Gammam_k^\star$ \eqref{eq:NNLS} corresponds to the $l_k$-th multi-path component in $\check{\sfH}_{k,s}(t,\tau)$. Consequently, the beamforming vector for the $k$-th UE at the BS is given by
\begin{align}\label{tx_MR}
\bfu_{k,1} =  \Fm_D\cdot \check{\bfu}_{k,1},
\end{align}
Further, the analog precoding vector support $\overline{\bfU}^{\text{BST}} $\eqref{RF_support}  and the baseband precoding matrix $\bfA_B^{\text{BST}}$ are given by $
\overline{\bfU}^{\text{BST}} = [\bfu_{1,1}, \bfu_{2,1}, ..., \bfu_{K,1}] $
and $\bfA_B^{\text{BST}}= \bfI_K$, respectively. In this case, an additional uplink channel estimation of $\widetilde{\sfH}_{s}(t,\tau)$ can be omitted. The eventual $D\times K$ BST precoder in \eqref{precoder} reads
\begin{align}
\bfU^{\text{BST}} =\overline{\bfU}^{\text{BST}} \!\!\cdot \bfA_B^{\text{BST}} = \overline{\bfU}^{\text{BST}}.
\end{align}

\subsubsection{Baseband Zeroforcing (BZF) Scheme}
In this scheme, we consider ZF precoding for potential multi-user interference cancellation. More precisely, we assume that after a BA phase as in Section \ref{beamAlignment},  each UE steers its beam towards the estimated strongest AoA $\bfv_{k} $ \eqref{receive_vector}. Meanwhile, each UE feeds back to the BS the AoD information of its top-$p$ strongest paths along the direction $\bfv_{k}= \Fm_N \check{\bfv}_{k}$, where $1\leq p\ll M$. Let $\bfr_k\in \bR_+^{M}$ denote the non-negative second-order channel statistics corresponding to the $k$-th UE, given by
\begin{align}
\bfr_{k} = ({\Gammam}_k^{\star})^\herm \cdot\check{\bfv}_{k},
\end{align}
where ${\Gammam}_k^\star$ is given in \eqref{eq:NNLS}. Let the elements in $\bfr_{k}$ be arranged in non-increasing order in terms of their strengths, i.e., $\bfr_{k}[m_1^k]\geq \bfr_{k}[m_2^k] \geq ... \bfr_{k}[m_p^k] \geq ... \geq \bfr_{k}[m_M^k]$, and define $\clA_k=\{m_1^k, m_2^k, ..., m_p^k\}$ as the beam index set of the top-$p$  strongest elements in $\bfr_{k}$. We assume that each UE feeds back its beam index set $\clA_k$ to the BS through the RACCH slots as illustrated in \figref{2_frame_timeline}. Consequently, the analog precoding vector support $\overline{\bfU}^{\text{ZF}}$ at the BS can be written as
\begin{align}\label{RF_supp_ZF}
\overline{\bfU}^{\text{ZF}} = [\bff_{D,m_1^1}, ...,\bff_{D,m_p^1}, ..., \bff_{D,m_1^K}, ...,  \bff_{D,m_p^K}], 
\end{align} 
where $\bff_{D,i}$ denotes the $i$-th column of the DFT matrix $\bfF_D$. Substituting \eqref{RF_supp_ZF} into \eqref{EF_H}, the effective channel $\widetilde{\sfH}_{s}(t,\tau)$ reads
\begin{align}
\widetilde{\sfH}_{s}(t,\tau)=\bfV^\herm \cdot \overline{\sfH}_s(t,\tau) \cdot \overline{\bfU}^{\text{ZF}},
\end{align}
which can be estimated through an ``exhaustive'' procedure with orthogonal uplink pilots at the cost of $(p\cdot K\ll MN)$ sub-slots. As a result, the baseband precoding matrix $\bfA_B^{\text{ZF}}$ can be written as
\begin{align}
\bfA_B^{\text{ZF}} = \widetilde{\sfH}_{s}(t,\tau)^\herm\cdot\left(\widetilde{\sfH}_{s}(t,\tau)\widetilde{\sfH}_{s}(t,\tau)^\herm\right)^{-1}\cdot \Delta^{\text{ZF}},
\end{align}
where $\Delta^{\text{ZF}}\in \bR_+^{K\times K}$ is a diagonal matrix, taking into account the total radiated power constraint. The eventual BZF precoder is then given by
\begin{align}
\bfU^{\text{ZF}} =\overline{\bfU}^{\text{ZF}} \!\!\cdot \bfA_B^{\text{ZF}}.
\end{align}

In the following section, we will compare the asymptotic sum spectral efficiency in terms of different transmitter architectures. To effectively capture the channel quality before BA,  we also define the SNR before beamforming (BBF) by
\begin{align}\label{snrBBF}
\snrbef =\frac{\ptot \sum_{l=1}^{L}\gamma_l}{N_0B}.
\end{align}
This is the SNR obtained when a single pilot stream ($M_{\text{RF}} = 1$) is transmitted through a single BS antenna and is received at a single UE antenna (isotropic transmission) via a single RF chain ($N_{\text{RF}} = 1$) over the whole bandwidth $B$. 

\section{Numerical Results}\label{Numerical}

We consider a system with a BS using $M=32$ antennas and $M_{\text{RF}}=2$ RF chains and each UE using $N=16$ antennas and $N_{\text{RF}}=1$ RF chain.
The system is assumed to work at $f_0=40$ GHz with a maximum available bandwidth of $B=0.8$ GHz. We assume the channel for each UE contains $L_k=3$ links given by $(\gamma_{k,1}=1,\eta_{k,1}=100)$, $(\gamma_{k,2}=0.6,\eta_{k,2}=10)$, and $(\gamma_{k,3}=0.6,\eta_{k,3}=0)$ as defined in \eqref{rice_fading}. In the following, we will compare the performance of the two transmitter architectures in \figref{TX}.


\subsection{Training Efficiency for the Initial Beam Alignment Phase}
As illustrated in \figref{1-BA-SE} (a), due to the fact that the OSPS architecture has lower angular resolution and encounters larger sidelobe power leakage than the FC case, the former requires moderately $\sim 20$ more beacon slots than the latter for $P_D\geq 0.95$. We also simulate a recent time-domain BA algorithm based on \cite{AlkhateebTimeDomain2017} which focuses on estimating the instantaneous channel coefficients with an  orthogonal matching pursuit (OMP) technique. As we can see, for both transmitter architectures the proposed BA scheme requires much less training overhead than that in \cite{AlkhateebTimeDomain2017}, implying its advantage for practical fast channel connection.


\begin{figure*}[t]
	\centering
	\hspace{-0.8cm}
	\begin{subfigure}[b]{0.3\textwidth}
		\centering
		\scalebox{.6}{\input{3-compare}}%
	\end{subfigure}
	\hspace{0.3cm}
	\begin{subfigure}[b]{0.3\textwidth}
		\centering
		\scalebox{.6}{\input{1-P-P}}%
	\end{subfigure}
	\hspace{0.3cm}
	\begin{subfigure}[b]{0.3\textwidth}
		\centering
		\scalebox{.6}{\input{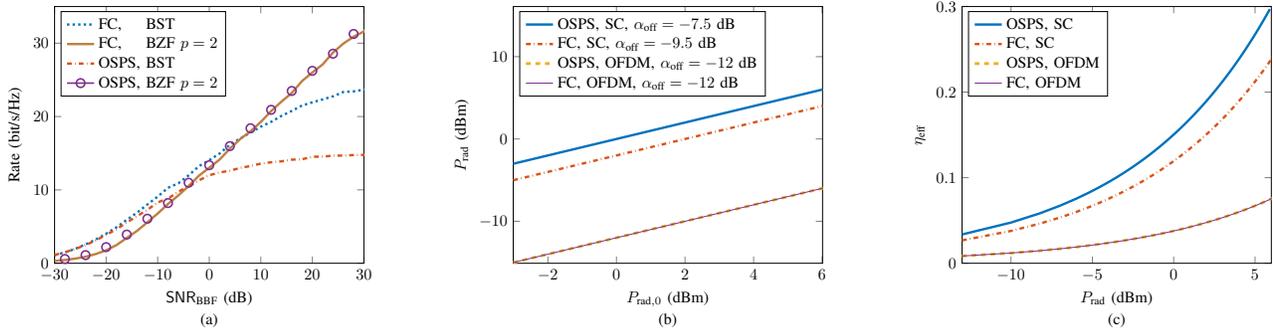}}%
	\end{subfigure}
	\caption{{\small The performance evaluation of different transmitter architectures in terms of (a) the sum spectral efficiency  vs. increasing $\snrbef$, (b) the actual radiated power under {\em Option I} vs. the radiated power of the reference scenario, (c) the power efficiency under {\em Option II} vs. the actual radiated power.}}
	\label{1-SE-PAE}
\end{figure*}

\subsection{Comparison of Different Precoding Schemes}
To first evaluate the efficiency of the proposed precoding schemes,  we simulate the sum spectral efficiency with $K=2$. As shown in \figref{1-BA-SE} (b), for the FC transmitter, in the range of $\snrbef\leq 10\,$dB the simple BST scheme achieves the highest sum spectral efficiency, whereas when $\snrbef\gg 0\,$dB the BZF precoder performs better. Also, the curves of BZF precoders with different $p$ values coincide with each other. Namely, the choice of $p$ plays a trade-off between mobility and the overhead for additional channel estimation but does not change the sum spectral efficiency. Further, the OSPS transmitter achieves similar performance. As a comparison, we also simulate a recent hybrid precoding scheme proposed in \cite{Castellanos2018} which is completely based on downlink channel reconstruction. As we can see, the proposed precoders achieve much better performance than that in \cite{Castellanos2018}.

\subsection{Fully-Connected or One-Stream-Per-Subarray?}
Note that the performance of different architectures highly depends on the channel condition ($\snrbef$) and the underlying precoders. A doubtless fact is that, the hardware complexity of \figref{TX} (b) is much lower than \figref{TX}(a). For the same channel condition, \figref{TX} (b) requires a slightly less initial training overhead. As for the data communication phase, given the parameters in this paper, we can see from \figref{1-SE-PAE} (a) that by using the BST precoder under weak channel conditions (i.e., $\snrbef<0\,$dB) and using the BZF precoder under strong channel conditions (i.e., $\snrbef\geq0\,$dB), the two architectures achieve a similar sum spectral efficiency.

To evaluate the architecture power efficiency, otherwise stated, we consider the BST precoder. We first assume a reference scenario as the baseline, i.e, the OSPS architecture using the BST precoder and a SC modulation, with PAs of $P_{\text{max},0}=6$ dBm, $\eta_{\text{max},0}=0.3$. The backoff factor with respect to different waveforms and transmitter architectures can be written as  $\alpha_{\text{off}}=1/(P_{\text{PAPR}})$, where $P_{\text{PAPR}}$ represents the PAPR of the input signals at the PAs. The investigation for 3GPP LTE in \cite{MyungPAPR} showed that with a probability of $0.9999$, the PAPR of the LTE SC waveform is smaller than $\sim7.5\,$dB and the PAPR of the LTE orthogonal frequency division multiplexing (OFDM) waveform (with $512$ subcarriers employing QPSK) is smaller than $\sim12\,$dB. We set $P_{\text{PAPR}}$ to these values for \figref{TX} (b). In \figref{TX} (a), however, the input signals of the PAs are the sum of the signals from different RF chains. For OFDM signaling each signal can be modeled as a Gaussian random process \cite{MyungPAPR} and the signals from different RF chains are independent, hence, the PAPR of the sum is the same as of one RF chain. For the case of SC signaling there is no clear work in the literature that shows how the sum of SC signals behaves. We simulated the sum of $M_{\text{RF}}=2$ SC signals using the same parameters as in \cite{MyungPAPR}. The result shows that with probability of 0.9999 the PAPR of the sum is smaller than $\sim9.5\,$dB. We apply these values and without loss of generality, we assume $\alpha_{\text{off},0} = -7.5\,$dB for the reference scenario. As shown in \eqref{eff1}, by deploying the same PAs ({\em Option I}), the two architectures achieve the same efficiency for a given $P_{\text{rad}}$. However as illustrated in \figref{1-SE-PAE} (b), given the same input signal (after the power compensation for the FC architecture) and precoding matrix as in the reference scenario, the 
OSPS architecture with SC signaling (OSPS, SC) achieves the highest $P_{\text{rad}}$, followed by (FC, SC), (OSPS, OFDM), and (FC, OFDM). In contrast, by deploying different PAs ({\em Option II})\footnote{Since $\eta_{\text{max}}$ of different PAs highly depends on the technology, for simplicity, we assume that different PAs working in their linear range have roughly the same maximum efficiency $\eta_{\text{max},0}$.}, \figref{1-SE-PAE} (c) shows that (OSPS, SC) achieves the highest power efficiency, followed by (FC, SC), (OSPS, OFDM) and (FC, OFDM).

\section{Conclusion}

In this paper, we proposed an analysis framework to evaluate the performance of typical hybrid transmitters at mmWave frequencies. In particular, we focused on the comparison of
a fully-connected (FC) architecture  and a one-stream-per-subarray (OSPS) architecture. 
We jointly evaluated the performance of the two architectures in terms of the initial  beam alignment (BA), the data communication, and the transmitter power efficiency. 
We used our recently proposed BA scheme and a simple precoding scheme based on zero-forcing precoding of the effective channel after BA. Both schemes 
outperform the state-of-the-art counterparts in the literature and can be considered as the de-facto new state of the art. 
Given the parameters in this paper, our simulation results show that the two architectures achieve a similar sum spectral efficiency, 
but the OSPS architecture outperforms the FC case in terms of hardware complexity and power efficiency, 
only at the cost of a slightly longer time for the initial BA. 
We hope that the proposed work provides a good analysis framework for future mmWave MU-MIMO system design.

\balance
{\footnotesize
	\bibliographystyle{IEEEtran}
	\bibliography{references}
}

\end{document}

%% file: symbols.tex
\usepackage{mathbbol}

\def\mindex#1{\index{#1}}

\def\sq{\hbox{\rlap{$\sqcap$}$\sqcup$}}
\def\qed{\ifmmode\sq\else{\unskip\nobreak\hfil
\penalty50\hskip1em\null\nobreak\hfil\sq
\parfillskip=0pt\finalhyphendemerits=0\endgraf}\fi\medskip}

\long\def\defbox#1{\framebox[.9\hsize][c]{\parbox{.85\hsize}{%
\parindent=0pt
\baselineskip=12pt plus .1pt      
\parskip=6pt plus 1.5pt minus 1pt 
 #1}}}

\long\def\beginbox#1\endbox{\subsection*{}%
\hbox{\hspace{.05\hsize}\defbox{\medskip#1\bigskip}}%
\subsection*{}}

\def\endbox{}

\def\diag{{\text{diag}}}

\newsavebox{\junk}
\savebox{\junk}[1.6mm]{\hbox{$|\!|\!|$}}

\def\argmin{\mathop{\rm arg\, min}}

\def\bC{{\mathbb C}}

\def\bE{{\mathbb E}}

\def\bR{{\mathbb R}}

\def\bfA{{\bf A}}
\def\bfB{{\bf B}}

\def\bfF{{\bf F}}

\def\bfI{{\bf I}}

\def\bfU{{\bf U}}
\def\bfV{{\bf V}}

\def\bfa{{\bf a}}

\def\bff{{\bf f}}

\def\bfq{{\bf q}}
\def\bfr{{\bf r}}

\def\bfu{{\bf u}}
\def\bfv{{\bf v}}
\def\bfw{{\bf w}}
\def\bfx{{\bf x}}
\def\bfy{{\bf y}}
\def\bfz{{\bf z}}

\def\tto{{\mathtt o}}

\def\ttt{{\mathtt t}}

\def\sfH{{\sf H}}

\def\bfmath#1{{\mathchoice{\mbox{\boldmath$#1$}}%
{\mbox{\boldmath$#1$}}%
{\mbox{\boldmath$\scriptstyle#1$}}%
{\mbox{\boldmath$\scriptscriptstyle#1$}}}}

\def\bfmY{\bfmath{Y}}

\def\bfmhhaY{\bfmath{\hhaY}} 
\def\bfmhhaY{\hbox to 0pt{$\widehat{\bfmY}$\hss}\widehat{\phantom{\raise 1.25pt\hbox{$\bfmY$}}}}

\def\til={{\widetilde =}}

\def\clA{{\cal A}}

\def\clC{{\cal C}}

\def\clN{{\cal N}}

 \def\FRAC#1#2#3{\genfrac{}{}{}{#1}{#2}{#3}}

\def\ddtp{{\mathchoice{\FRAC{1}{d^{\hbox to 2pt{\rm\tiny +\hss}}}{dt}}%
{\FRAC{1}{d^{\hbox to 2pt{\rm\tiny +\hss}}}{dt}}%
{\FRAC{3}{d^{\hbox to 2pt{\rm\tiny +\hss}}}{dt}}%
{\FRAC{3}{d^{\hbox to 2pt{\rm\tiny +\hss}}}{dt}}}}

\def\average#1,#2,{{1\over #2} \sum_{#1}^{#2}}

\def\eye(#1){{\bf(#1)}\quad}

\def\eq#1/{(\ref{e:#1})}

\newcommand{\beqn}[1]{\notes{#1}%
\begin{eqnarray} \elabel{#1}}

\newcommand{\eeqn}{\end{eqnarray} }

\newcommand{\beq}[1]{\notes{#1}%
\begin{equation}\elabel{#1}}

\newcommand{\eeq}{\end{equation}}

\def\bdes{\begin{description}}
\def\edes{\end{description}}

\newcounter{rmnum}

\newcounter{anum}

{\end{list}}

\def\ass(#1:#2){(#1\ref{#1:#2})}

\def\ritem#1{
\item[{\sf \ass(\current_model:#1)}]
}

\newenvironment{recall-ass}[1]{%
\begin{description}
\def\current_model{#1}}{
\end{description}
}

\def\ptot{{P_{\ttt \tto \ttt}}}

%% file: macros.tex
\long\def\comment#1{}

\newcommand{\Fm}{{\bf F}}

\newcommand{\Gammam}{\boldsymbol{\Gamma}}

\newcommand{\trace}{{\hbox{tr}}}

\newcommand{\transp}{{\sf T}}
\renewcommand{\vec}{{\rm vec}}



%% file: 1-BA-compare.tex
%
%
\definecolor{mycolor1}{rgb}{0.00000,0.44700,0.74100}%
\definecolor{mycolor2}{rgb}{0.85000,0.32500,0.09800}%
%


\begin{tikzpicture}

\begin{axis}[%
xmin=10,
xmax=100,
xlabel={Number of beacon slots $T$},
ymin=0.2,
ymax=1.02,
ytick={  0, 0.2, 0.4, 0.6, 0.8,   1},
yminorticks=true,
ylabel={$P_D$},
title style={at={(0.5,0)},anchor=north,yshift=-1.2cm},
title = {(a)},
legend style={at={(0.48,0.01)},nodes={scale=1, transform shape}, anchor=south west, legend cell align=left, align=left, draw=black}
]
\addplot [color=mycolor1, line width=1.5pt]
  table[row sep=crcr]{%
5	0.055\\
10	0.1775\\
15	0.3\\
20	0.45\\
25	0.6\\
30	0.78\\
35	0.88\\
40	0.935\\
45	0.975\\
50	0.99\\
55	0.99\\
60	0.995\\
65	0.995\\
70	1\\
75	1\\
80	1\\
85	1\\
90	1\\
95	1\\
100	1\\
};
\addlegendentry{fully, NNLS}

\addplot [color=mycolor2, dashed, line width=1.5pt]
  table[row sep=crcr]{%
5	0.035\\
10	0.2\\
15	0.285\\
20	0.347781643521727\\
25	0.42167246528259\\
30	0.495563287043453\\
35	0.565\\
40	0.6775\\
45	0.79\\
50	0.84\\
55	0.91\\
60	0.945\\
65	0.965\\
70	0.975\\
75	0.995\\
80	0.994999998139344\\
85	0.995\\
90	0.995000000356496\\
95	0.994999999929652\\
100	0.994999999367972\\
105	0.995000001061258\\
110	0.99500000032429\\
115	0.995\\
120	0.995\\
125	0.995\\
130	0.995\\
135	0.995\\
140	0.995\\
145	0.995\\
150	1\\
};
\addlegendentry{sub,$\,\,$ NNLS}

\addplot [color=black, dashdotted, line width=1.5pt]
table[row sep=crcr]{%
5	0\\
15	0.1\\
25	0.25\\
35	0.4\\
45	0.5\\
55	0.6\\
65	0.64914781400408\\
75	0.72\\
85	0.8\\
95	0.846\\
100	0.86\\
};
\addlegendentry{fully, OMP \cite{AlkhateebTimeDomain2017}}

\end{axis}
\end{tikzpicture}%

%% file: 1-fully-rate-snr.tex
%
%
\definecolor{mycolor1}{rgb}{0.00000,0.44700,0.74100}%
\definecolor{mycolor2}{rgb}{0.85000,0.32500,0.09800}%
\definecolor{mycolor3}{rgb}{0.92900,0.69400,0.12500}%
\definecolor{mycolor4}{rgb}{0.49400,0.18400,0.55600}%
\definecolor{mycolor5}{rgb}{0.46600,0.67400,0.18800}%
\definecolor{mycolor6}{rgb}{0.30100,0.74500,0.93300}%
%


\begin{tikzpicture}

\begin{axis}[%
xmin=-30,
xmax=30,
xlabel={$\snrbef$ (dB)},
ymin=0.0,
ymax=40,
ytick={  0, 10,20,30,40},
yminorticks=true,
ylabel={Rate (bit/s/Hz)},
title style={at={(0.5,0)},anchor=north,yshift=-1.2cm},
title = {(b)},
legend style={at={(0.02,0.58)},nodes={scale=1, transform shape}, anchor=south west, legend cell align=left, align=left, draw=black}
]
\addplot [color=mycolor1, dotted, line width=1.5pt]
  table[row sep=crcr]{%
-30	1.10888546385037\\
-28	1.48337872462583\\
-26	1.96838339096641\\
-24	2.56281643947765\\
-22	3.23757446653042\\
-20	4.03602540037747\\
-18	4.90345797808626\\
-16	5.91628029394328\\
-14	6.9006268037376\\
-12	8.00811702784691\\
-10	9.08781153369478\\
-8	10.2892144078029\\
-6	10.9001333749274\\
-4	11.9784878753188\\
-2	13.2329495586026\\
0	14.0554715695444\\
2	14.9507945610856\\
4	16.1539952652149\\
6	17.0026811937513\\
8	17.8050534843391\\
10	18.6152108753653\\
12	19.276393292642\\
14	20.0837301669065\\
16	20.7634495314291\\
18	21.5023999768866\\
20	21.9398083593496\\
22	22.3315772398021\\
24	22.7495350481057\\
26	23.3430250279165\\
28	23.4302434242649\\
30	23.6613300093647\\
32	23.8446489722868\\
34	24.0056782642066\\
36	24.4726649733355\\
38	24.5983801653085\\
40	24.9306448812021\\
};
\addlegendentry{FC, BST}

\addplot [color=mycolor2, dashed, line width=1.5pt]
  table[row sep=crcr]{%
-30	0.443463512794568\\
-28	0.596270887605884\\
-26	0.816554314138728\\
-24	1.09048398061888\\
-22	1.45963151971679\\
-20	1.95434915790041\\
-18	2.62516905570737\\
-16	3.46676177090052\\
-14	4.36722871665759\\
-12	5.47310038725401\\
-10	6.57742002904563\\
-8	7.85150912809715\\
-6	9.01046974498231\\
-4	10.2586186490776\\
-2	11.6891080179239\\
0	12.7753645151945\\
2	13.8846331363419\\
4	15.3840158339368\\
6	16.6798276137438\\
8	17.743380800955\\
10	18.9978494805004\\
12	20.4888106845969\\
14	21.7924967687615\\
16	22.9660261345213\\
18	24.5827927843714\\
20	25.7582183378936\\
22	26.85771560546\\
24	28.4570225732075\\
26	29.703566535389\\
28	30.6206800187042\\
30	31.3677089562977\\
32	33.0523219783871\\
34	34.6816007386283\\
36	35.9858158161102\\
38	37.0868556435283\\
40	39.7975235344294\\
};
\addlegendentry{FC, BZF $p=1$}

\addplot [color=mycolor4, draw=none, line width = 1pt, mark=o, mark size = 2.7pt, mark options={solid, mycolor4}]
  table[row sep=crcr]{%
-28	0.441493647103472\\
-24	0.885674845578455\\
-20	1.76192475951479\\
-16	3.44964854140042\\
-12	5.58489067820643\\
-8	8.07861879332269\\
-4	10.5147173549389\\
0	13.0147416331256\\
4	15.618218510393\\
8	17.9807321284435\\
12	20.7327982519816\\
16	23.2008523926016\\
20	26.0018435552277\\
24	28.7135409139831\\
28	30.8779229716275\\
32	33.2722131568113\\
36	36.2277766186696\\
40	38.9965346458743\\
};
\addlegendentry{FC, BZF $p=2$}

\addplot [color=brown, line width=1.0pt]
  table[row sep=crcr]{%
-30	0.255997882629152\\
-28	0.359388677440939\\
-26	0.530946174512797\\
-24	0.760769467043944\\
-22	1.1190901741029\\
-20	1.60098719826192\\
-18	2.32344960439049\\
-16	3.29397790124603\\
-14	4.3192717576348\\
-12	5.52967198211997\\
-10	6.77670862101993\\
-8	8.11292004261698\\
-6	9.28632942290635\\
-4	10.580331982279\\
-2	12.0216578228826\\
0	13.0958053174434\\
2	14.209145756089\\
4	15.7011787847697\\
6	17.0079808588118\\
8	18.0543190266503\\
10	19.3335931062133\\
12	20.8160708506233\\
14	22.1167793528023\\
16	23.2772580105707\\
18	24.9184171397827\\
20	26.0843402110575\\
22	27.1815305535484\\
24	28.7961469585357\\
26	30.008501869864\\
28	30.9550255517737\\
30	31.6821885414649\\
32	33.357109727905\\
34	35.0430142773228\\
36	36.3014343928005\\
38	37.3614504422415\\
40	39.0747444207643\\
};
\addlegendentry{FC, BZF $p=3$}


\addplot [color=black, line width=1.5pt, dashdotted]
  table[row sep=crcr]{%
-30	0.441808991303706\\
-26	0.716457145719879\\
-22	1.07384445846245\\
-18	1.55410088994535\\
-14	2.25913846248107\\
-10	3.37358909654144\\
-6	5.21498004765062\\
-2	7.65094290184979\\
2	11.1040378743663\\
6	14.490522626308\\
10	18.1422375688022\\
14	21.6180401625736\\
18	24.3567192365815\\
22	27.1887827784319\\
26	29.3498592069184\\
30	31.2131088552243\\
34	32.5485437428585\\
38	33.6949460175948\\
};
\addlegendentry{FC, precoder in \cite{Castellanos2018}}

\end{axis}
\end{tikzpicture}%
%

%% file: 2-sub-rate-snr.tex
%
%
\definecolor{mycolor1}{rgb}{0.00000,0.44700,0.74100}%
\definecolor{mycolor2}{rgb}{0.85000,0.32500,0.09800}%
\definecolor{mycolor3}{rgb}{0.92900,0.69400,0.12500}%
\definecolor{mycolor4}{rgb}{0.49400,0.18400,0.55600}%
\definecolor{mycolor5}{rgb}{0.46600,0.67400,0.18800}%
\definecolor{mycolor6}{rgb}{0.30100,0.74500,0.93300}%
%


\begin{tikzpicture}

\begin{axis}[%
xmin=-30,
xmax=30,
xlabel={$\snrbef$ (dB)},
ymin=0.0,
ymax=40,
ytick={  0, 10,20,30,40},
yminorticks=true,
ylabel={Rate (bit/s/Hz)},
title style={at={(0.5,0)},anchor=north,yshift=-1.2cm},
title = {(c)},
legend style={at={(0.02,0.58)},nodes={scale=1, transform shape}, anchor=south west, legend cell align=left, align=left, draw=black}
]
\addplot [color=mycolor1, dotted, line width=1.5pt]
  table[row sep=crcr]{%
-30	1.13014187799955\\
-28	1.48454092599332\\
-26	1.95582776702461\\
-24	2.47175174422937\\
-22	3.12444767426163\\
-20	3.88096337730023\\
-18	4.62960445545426\\
-16	5.55733724558078\\
-14	6.42773916895651\\
-12	7.27839124508613\\
-10	8.3023118768703\\
-8	8.80603243251427\\
-6	9.76384646059467\\
-4	10.492626152101\\
-2	11.3006627458648\\
0	11.9957142960766\\
2	12.3796189349761\\
4	12.613247360457\\
6	12.9953380685527\\
8	13.2531706764938\\
10	13.5852679892223\\
12	13.7628165628072\\
14	13.8924703077583\\
16	14.1209086423472\\
18	14.1730537836225\\
20	14.5184325957209\\
22	14.5322960243303\\
24	14.6685561614562\\
26	14.6752897584317\\
28	14.7510363669014\\
30	14.7774340045138\\
32	15.0165161876645\\
34	15.072807137961\\
36	15.1970561080271\\
38	15.3231270250322\\
40	15.433330654718\\
};
\addlegendentry{OSPS, BST}

\addplot [color=mycolor2, dashed, line width=1.5pt]
  table[row sep=crcr]{%
-30	0.471086121229096\\
-28	0.629672108775695\\
-26	0.845360162459799\\
-24	1.10444249555705\\
-22	1.48074009672012\\
-20	2.04196859263841\\
-18	2.6348770281309\\
-16	3.53123773567794\\
-14	4.519810766007\\
-12	5.61254258446743\\
-10	6.78126463617967\\
-8	7.76905659342017\\
-6	9.11857994220859\\
-4	10.5365337364073\\
-2	11.6569093816683\\
0	12.9202088095647\\
2	14.2168222392209\\
4	15.5164775479177\\
6	16.8161521443403\\
8	17.9625192048739\\
10	19.110524197741\\
12	20.4613028075398\\
14	21.8506519794017\\
16	23.0206512294232\\
18	24.4513107637632\\
20	25.7687953726457\\
22	27.2451120717163\\
24	28.099064678231\\
26	29.2983091247035\\
28	30.8093071246621\\
30	32.1944420179069\\
32	33.3827243013519\\
34	35.0080571825432\\
36	36.1850701652494\\
38	37.1864207783344\\
40	39.8412390185542\\
};
\addlegendentry{OSPS, BZF $p=1$}

\addplot [color=mycolor4, draw=none, line width = 1pt, mark=o, mark size = 2.7pt, mark options={solid, mycolor4}]
table[row sep=crcr]{%
-28	0.579130093143652\\
-24	1.10964941480869\\
-20	2.19886141101585\\
-16	3.90906767045824\\
-12	6.06986299086299\\
-8	8.206912632557\\
-4	10.9784251688571\\
0	13.3651678365376\\
4	15.9661971645583\\
8	18.403387911029\\
12	20.912462457225\\
16	23.4871723164321\\
20	26.2267490521353\\
24	28.5672995253482\\
28	31.2562400840611\\
32	33.8469338592643\\
36	36.6505046838948\\
40	39.3409841336341\\
};
\addlegendentry{OSPS, BZF $p=2$}

\addplot [color=brown, line width=1.0pt]
  table[row sep=crcr]{%
-30	0.386197455536119\\
-28	0.550275004074319\\
-26	0.789933655520851\\
-24	1.10763865150837\\
-22	1.58445099299062\\
-20	2.29306119366199\\
-18	3.04783391866397\\
-16	4.11500247307055\\
-14	5.21525535867814\\
-12	6.37436297133916\\
-10	7.5695778365386\\
-8	8.52990647807733\\
-6	9.90048848587214\\
-4	11.3297169933466\\
-2	12.4604294440077\\
0	13.7005435983412\\
2	14.9993738410158\\
4	16.3081386485522\\
6	17.6060973269511\\
8	18.7415498077125\\
10	19.9104479619869\\
12	21.2425889894978\\
14	22.6332877459004\\
16	23.8202126080853\\
18	25.2350727634455\\
20	26.5713150500237\\
22	28.0325929892939\\
24	28.8875440016936\\
26	30.067073021605\\
28	31.5951387159408\\
30	32.9975259402678\\
32	34.1824035888199\\
34	35.8139707666247\\
36	36.9733117528282\\
38	37.9157046646964\\
40	39.6786802276034\\
};
\addlegendentry{OSPS, BZF $p=3$}


\addplot [color=black, line width=1.5pt, dashdotted]
  table[row sep=crcr]{%
-30	0.447937070390076\\
-28	0.58282126193076\\
-26	0.735733929046077\\
-24	0.92053974126235\\
-22	1.12841478413959\\
-20	1.39251412082889\\
-18	1.73347479214506\\
-16	2.16268306030866\\
-14	2.7139896167145\\
-12	3.3929734421656\\
-10	4.20434284827691\\
-8	5.20689521976486\\
-6	6.41822896768735\\
-4	7.93192848466678\\
-2	9.33271513833391\\
0	10.9473856474713\\
2	12.6645172877747\\
4	14.6014368658173\\
6	15.9271271560815\\
8	17.49663850265\\
10	19.0694909063841\\
12	20.2928945500558\\
14	21.8672444984991\\
16	22.7557089275902\\
18	23.860396183876\\
20	25.0717699166148\\
22	25.7925805763898\\
24	26.6946365481986\\
26	27.7478516006314\\
28	28.382813401951\\
30	28.9682711623929\\
32	29.5550614513966\\
34	30.151088753272\\
36	30.4423168058538\\
38	30.9136927529167\\
40	31.3044841511617\\
};
\addlegendentry{OSPS, precoder in \cite{Castellanos2018}}

\end{axis}
\end{tikzpicture}%


%% file: 3-compare.tex
%
%
\definecolor{mycolor1}{rgb}{0.00000,0.44700,0.74100}%
\definecolor{mycolor2}{rgb}{0.85000,0.32500,0.09800}%
\definecolor{mycolor3}{rgb}{0.92900,0.69400,0.12500}%
\definecolor{mycolor4}{rgb}{0.49400,0.18400,0.55600}%
%

\begin{tikzpicture}

\begin{axis}[%
xmin=-30,
xmax=30,
xlabel={$\snrbef$ (dB)},
ymin=0.0,
ymax=35,
ytick={  0, 10,20,30,40},
yminorticks=true,
ylabel={Rate (bit/s/Hz)},
title style={at={(0.5,0)},anchor=north,yshift=-1.2cm},
title = {(a)},
legend style={at={(0.02,0.65)},nodes={scale=1, transform shape}, anchor=south west, legend cell align=left, align=left, draw=black}
]
\addplot [color=mycolor1, dotted, line width=1.5pt]
  table[row sep=crcr]{%
-30	1.10888546385037\\
-28	1.48337872462583\\
-26	1.96838339096641\\
-24	2.56281643947765\\
-22	3.23757446653042\\
-20	4.03602540037747\\
-18	4.90345797808626\\
-16	5.91628029394328\\
-14	6.9006268037376\\
-12	8.00811702784691\\
-10	9.08781153369478\\
-8	10.2892144078029\\
-6	10.9001333749274\\
-4	11.9784878753188\\
-2	13.2329495586026\\
0	14.0554715695444\\
2	14.9507945610856\\
4	16.1539952652149\\
6	17.0026811937513\\
8	17.8050534843391\\
10	18.6152108753653\\
12	19.276393292642\\
14	20.0837301669065\\
16	20.7634495314291\\
18	21.5023999768866\\
20	21.9398083593496\\
22	22.3315772398021\\
24	22.7495350481057\\
26	23.3430250279165\\
28	23.4302434242649\\
30	23.6613300093647\\
32	23.8446489722868\\
34	24.0056782642066\\
36	24.4726649733355\\
38	24.5983801653085\\
40	24.9306448812021\\
};
\addlegendentry{FC,$\,\,\,\,\,\,\,$ BST}

\addplot [color=brown, line width=1.5pt]
  table[row sep=crcr]{%
-30	0.319528198662643\\
-28	0.441493647103472\\
-26	0.634777906949563\\
-24	0.885674845578455\\
-22	1.26392171751523\\
-20	1.76192475951479\\
-18	2.47871700755387\\
-16	3.44964854140042\\
-14	4.41714466666424\\
-12	5.58489067820643\\
-10	6.77924808358353\\
-8	8.07861879332269\\
-6	9.24251604925386\\
-4	10.5147173549389\\
-2	11.9398091062369\\
0	13.0147416331256\\
2	14.129946553667\\
4	15.618218510393\\
6	16.92407129933\\
8	17.9807321284435\\
10	19.2564605486353\\
12	20.7327982519816\\
14	22.0326859692212\\
16	23.2008523926016\\
18	24.8403737125647\\
20	26.0018435552277\\
22	27.1045976011649\\
24	28.7135409139831\\
26	29.9354281663936\\
28	30.8779229716275\\
30	31.6007724407068\\
32	33.2722131568113\\
34	34.9588354696036\\
36	36.2277766186696\\
38	37.2862578975093\\
40	38.9965346458743\\
};
\addlegendentry{FC,$\,\,\,\,\,\,\,$ BZF $p=2$}

\addplot [color=mycolor2, dashdotted, line width=1.5pt]
  table[row sep=crcr]{%
-30	1.13014187799955\\
-28	1.48454092599332\\
-26	1.95582776702461\\
-24	2.47175174422937\\
-22	3.12444767426163\\
-20	3.88096337730023\\
-18	4.62960445545426\\
-16	5.55733724558078\\
-14	6.42773916895651\\
-12	7.27839124508613\\
-10	8.3023118768703\\
-8	8.80603243251427\\
-6	9.76384646059467\\
-4	10.492626152101\\
-2	11.3006627458648\\
0	11.9957142960766\\
2	12.3796189349761\\
4	12.613247360457\\
6	12.9953380685527\\
8	13.2531706764938\\
10	13.5852679892223\\
12	13.7628165628072\\
14	13.8924703077583\\
16	14.1209086423472\\
18	14.1730537836225\\
20	14.5184325957209\\
22	14.5322960243303\\
24	14.6685561614562\\
26	14.6752897584317\\
28	14.7510363669014\\
30	14.7774340045138\\
32	15.0165161876645\\
34	15.072807137961\\
36	15.1970561080271\\
38	15.3231270250322\\
40	15.433330654718\\
};
\addlegendentry{OSPS, BST}

\addplot [color=mycolor4, line width=1.0pt, draw=none, mark=o, mark size =2.7pt, mark options={solid, mycolor4}]
  table[row sep=crcr]{%
-28	0.579130093143652\\
-24	1.10964941480869\\
-20	2.19886141101585\\
-16	3.90906767045824\\
-12	6.06986299086299\\
-8	8.206912632557\\
-4	10.9784251688571\\
0	13.3651678365376\\
4	15.9661971645583\\
8	18.403387911029\\
12	20.912462457225\\
16	23.4871723164321\\
20	26.2267490521353\\
24	28.5672995253482\\
28	31.2562400840611\\
32	33.8469338592643\\
36	36.6505046838948\\
40	39.3409841336341\\
};
\addlegendentry{OSPS, BZF $p=2$}

\end{axis}
\end{tikzpicture}%

%% file: 1-P-P.tex
%
%
\definecolor{mycolor1}{rgb}{0.00000,0.44700,0.74100}%
\definecolor{mycolor2}{rgb}{0.85000,0.32500,0.09800}%
\definecolor{mycolor3}{rgb}{0.92900,0.69400,0.12500}%
\definecolor{mycolor4}{rgb}{0.49400,0.18400,0.55600}%
\begin{tikzpicture}

\begin{axis}[%
xmin=-3.01,
xmax=6.01,
xlabel={$P_{\text{rad},0}$ (dBm)},
ymax=16.1,
ymin=-15.1,
yminorticks=true,
ylabel={$P_{\text{rad}}$ (dBm)},
title style={at={(0.5,0)},anchor=north,yshift=-1.2cm},
title = {(b)},
legend style={at={(0.02,0.65)},nodes={scale=1, transform shape}, anchor=south west, legend cell align=left, align=left, draw=black}
]
\addplot [color=mycolor1, line width=1.5pt]
  table[row sep=crcr]{%
-3.01029995663981	-3.01029995663981\\
0	0\\
1.76091259055681	1.76091259055681\\
3.01029995663981	3.01029995663981\\
3.97940008672038	3.97940008672038\\
4.77121254719662	4.77121254719662\\
5.44068044350276	5.44068044350276\\
6.02059991327962	6.02059991327962\\
};
\addlegendentry{OSPS, SC, $\alpha_{\text{off}}=-7.5$ dB}

\addplot [color=mycolor2, dashdotted, line width=1.5pt]
  table[row sep=crcr]{%
-3.01029995663981	-5.01029995663981\\
0	-2\\
1.76091259055681	-0.239087409443188\\
3.01029995663981	1.01029995663981\\
3.97940008672038	1.97940008672038\\
4.77121254719662	2.77121254719662\\
5.44068044350276	3.44068044350276\\
6.02059991327962	4.02059991327962\\
};
\addlegendentry{FC, SC, $\alpha_{\text{off}}=-9.5$ dB}

\addplot [color=mycolor3, dashed, line width=1.8pt]
  table[row sep=crcr]{%
-3.01029995663981	-15.0102999566398\\
0	-12\\
1.76091259055681	-10.2390874094432\\
3.01029995663981	-8.98970004336019\\
3.97940008672038	-8.02059991327962\\
4.77121254719662	-7.22878745280338\\
5.44068044350276	-6.55931955649724\\
6.02059991327962	-5.97940008672038\\
};
\addlegendentry{OSPS, OFDM, $\alpha_{\text{off}}=-12$ dB}

\addplot [color=mycolor4, line width=0.5pt]
  table[row sep=crcr]{%
-3.01029995663981	-15.0102999566398\\
0	-12\\
1.76091259055681	-10.2390874094432\\
3.01029995663981	-8.98970004336019\\
3.97940008672038	-8.02059991327962\\
4.77121254719662	-7.22878745280338\\
5.44068044350276	-6.55931955649724\\
6.02059991327962	-5.97940008672038\\
};
\addlegendentry{FC, OFDM, $\alpha_{\text{off}}=-12$ dB}

\end{axis}
\end{tikzpicture}%